\documentclass[a4paper,fleqn,useAMS,usenatbib]{mnras}

\usepackage{graphicx}
\usepackage{amsmath}
\usepackage{amssymb}
\usepackage{color}

\usepackage[T1]{fontenc}
\usepackage{ae,aecompl}

\usepackage{times}

\usepackage{multirow}

\title[Tracer bias around voids]{On the linearity of tracer bias around voids}

\author[G.~Pollina et al.]
{\parbox{\textwidth}{Giorgia Pollina$^{1, \, 2, }$\thanks{\href{mailto:gpollina@usm.lmu.de}{gpollina@usm.lmu.de}}, 
Nico Hamaus$^{1}$, 
Klaus Dolag$^{1, \, 3}$, 
Jochen Weller$^{1, \, 2, \, 4}$, 
Marco Baldi$^{5, \, 6, \, 7}$, 
Lauro Moscardini$^{5, \, 6, \, 7}$}
\\
\\$^{1}$Universit\"ats-Sternwarte M\"unchen, Fakult\"at f\"ur Physik, Ludwig-Maximilians Universit\"at M\"unchen, Scheinerstr. 1, D-81679 M\"unchen, Germany
\\$^{2}$Excellence Cluster Universe,  Boltzmannstr. 2, D-85748 Garching, Germany
\\$^{3}$Max-Planck-Institute for Astrophysics, Karl-Schwarzschild Strasse 1, D-85748 Garching, Germany
\\$^{4}$Max Planck Institute for Extraterrestrial Physics, Giessenbachstr. 1, D-85748 Garching, Germany
\\$^{5}$Dipartimento di Fisica e Astronomia, Alma Mater Studiorum Universit\`a di Bologna, viale Berti Pichat, 6/2, I-40127 Bologna, Italy
\\$^{6}$INAF - Osservatorio Astronomico di Bologna, via Ranzani 1, I-40127 Bologna, Italy
\\$^{7}$INFN - Sezione di Bologna, viale Berti Pichat 6/2, I-40127 Bologna, Italy
}

\date{Accepted 2017 March 28. Received 2017 March 28; in original form 2016 October 20}

\pubyear{2017}
\volume{469}

\begin{document}
\label{firstpage}
\pagerange{787--799}
\maketitle

\begin{abstract}
The large-scale structure of the universe can only be observed via luminous tracers 
of the dark matter. However, the clustering statistics of tracers are biased and depend 
on various properties, such as their host-halo mass and assembly history. On very large 
scales this tracer bias results in a constant offset in the clustering amplitude, known as \emph{linear bias}. 
Towards smaller non-linear scales, this is no longer the case and tracer bias becomes a complicated 
function of scale and time. We focus on tracer bias centred on cosmic voids, depressions of the 
density field that spatially dominate the universe. We consider three types of tracers: galaxies, 
galaxy clusters and AGN, extracted from the hydrodynamical simulation {\it Magneticum Pathfinder}. 
In contrast to common clustering statistics that focus on auto-correlations of tracers, we find that void-tracer 
cross-correlations are successfully described by a linear-bias relation. The tracer-density profile of voids 
can thus be related to their matter-density profile by a single number. We show that it coincides with the 
linear tracer bias extracted from the large-scale auto-correlation function and expectations from theory, 
if sufficiently large voids are considered. For smaller voids we observe a shift towards higher values. 
This has important consequences on cosmological parameter inference, as the problem of unknown 
tracer bias is alleviated up to a constant number. The smallest scales in existing datasets become 
accessible to simpler models, providing numerous modes of the density field that have been 
disregarded so far, but may help to further reduce statistical errors in constraining cosmology.
\end{abstract}






\section{Introduction}
\label{i}

In the present standard cosmological model the large-scale structure of the Universe
forms in a hierarchical process that begins with the gravitational collapse of 
overdense fluctuations of the matter density field into virialised and gravitationally
bound objects, known as dark matter haloes. Such objects provide the potential wells 
in which baryons can cool and condense to create galaxies that are now observed in the
sky \citep{peebles1980}. The understanding of modern cosmology and structure formation
is thereby deeply connected to the statistical properties of dark matter haloes and their 
hosted galaxies, which represent the final stage of the evolution of primordial fluctuations
and can be directly observed and used to constrain theory. Studying
the clustering properties of galaxies, it was discovered that they do not precisely 
mirror the clustering of the bulk of the dark matter distribution: such evidence brought
\citet{kaiser1984} to introduce the concept of galaxy bias to indicate that galaxies are
biased tracers of the underlying matter density field. \citet{kaiser1984} showed that
clusters of galaxies would naturally have a large bias, being rare objects that grow 
from the highest density peaks in the mass distribution.
Bias is now a known property of luminous tracers on very large scales, where density 
fluctuations are within the linear regime: in this case tracer bias is a simple constant 
offset in the clustering amplitude, known as {\it linear bias}. Towards small
scales, this elementary relation does not stand and bias becomes an unestablished 
function of scale and time.

In this paper we focus on the bias of tracers inside and around cosmic voids,
large underdense regions in the large-scale structure of the Universe 
that together with clusters, filaments and walls define the topology of the cosmic web 
as predicted in a cold dark matter (CDM) cosmology \citep{bond1996,pogosyan1998}. 
Voids are another peculiarity of the large scale structure of the Universe, representing
the result of the evolution of underdensities in the primordial density field.
Although the existence of voids has been one of the earliest predictions of the 
standard cosmological model \citep{hausman1983}, and the observational discovery of 
voids dates back to almost 40 years ago \citep[see][]{GregoryThompson1978, kirshner1981}, 
systematic studies about voids have become possible only recently. This is thanks to 
the increasing depth and volume of current galaxy surveys which map out larger and 
larger portions of the sky, and to the
advent of large cosmological simulations that are now capable to predict the 
distribution of matter in the cosmic web with very high accuracy.

The growing interest for cosmic voids in the literature is partly due to their not yet 
fully explored potential to constrain cosmology. Voids are the largest structures in 
the Universe and occupy most of its volume \citep{falck2015, cautun2014}; their 
spherically averaged density profile exhibits a universal shape that can be described 
by a simple empirical function \citep[see e.g.][]{colberg2005, hamaus2014, 
ricciardelli2013, ricciardelli2014}. According to the cosmological principle, they 
represent a population of statistically ideal spheres with a homogeneous distribution 
in the Universe at different redshifts, so that their observed shape evolution can be 
used to probe the expansion history of the Universe by means of the Alcock \& Paczynski 
(AP) test \citep{AlcockPaczynski1979}, as already demonstrated by recent works 
\citep{lavaux2012, sutter2012, sutter2014, hamaus2014b, hamaus2016, mao2016}. 

Void counts have the potential to improve upon current constraints on dark energy 
\citep{pisani2015} and can provide a test to discriminate between competing 
cosmological models. In fact, as their ordinary matter content is by definition 
very low, voids are expected to be most sensitive to the nature of Dark Energy 
(DE) and the features of the primordial density field in which they grow 
\citep{odrzywolek2009, damico2011,bos2012,gibbons2014}. It has been argued that 
the shape of voids is particularly sensitive to the equation of state of the DE 
component \citep{lavaux2010}. In addition, void abundance and void lensing are 
possible probes to test for modifications of gravity \citep{clampitt2013, Li_2011, 
Cai_etal_2014, Barreira_etal_2015, zivick2015}, couplings between dark matter 
and dark energy \citep{pollina2015}, the impact of massive neutrinos on structure 
formation \citep{massara2015} and possible degeneracies
between modifications of gravity and warm dark matter particle candidates \citep{baldi2016}. 
Although still controversial, it has been investigated 
whether the observed Cold Spot in the Cosmic Microwave Background (CMB) could be 
explained as Integrated Sachs-Wolfe (ISW) imprint caused by very large voids along 
the line of sight \citep[see e.g.][]{rees1968, Finelli_etal_2014, kovac2014, 
Nadathur2015coldspot} and in general, the potential of 
the ISW effect in voids is still under investigation \citep{kovacs2015, 
kovacsBellido2015, naidoo2016}. Finally, the clustering statistics of voids open up 
new opportunities to study cosmology \citep{hamaus2014V-gC, chan2014, clampitt2016, liang2016}. 
For example, the Baryon Acoustic Oscillation (BAO) feature has recently been detected in the auto-correlation 
function of voids \citep{kitakura2016}, providing a standard ruler from the underdense regime of the Universe.

All of these studies suggest voids can be considered as promising cosmological probes. 
Despite that, a lack of understanding in how to link void properties to theory, 
simulations and observations persists. For what concerns the theoretical comprehension 
of voids, one of the pioneering works in the field is presented in \citet{sheth2004}, 
where the authors provide a theory to model the void-size distribution and its evolution 
assuming spherical initial conditions, as commonly done in void evolution models 
\citep{peebles1980, blumenthal1992}. However, assuming voids to start evolving from 
spherical under-densities might not be representative of objects developing from 
Gaussian underdense fluctuations of arbitrary shape. The number function of voids 
identified in cosmological simulations is in fact not well represented by the model 
proposed by  \citet{sheth2004}, 
as recently argued (see e.g \citealt{jennings2013, nadathur2015, falck2015};
but see \citealt{pisani2015} for how to take this into account). Many studies have been 
conducted to better understand the evolution of voids over cosmic time \citep{achitouv2015, 
demchenko2016, wojtak2016} and their number function \citep{pycke2016}.

Similarly, another gap that still has to be bridged concerns the relation between 
the properties of voids in simulations with potentially observable voids. Numerous
catalogues of voids identified in spectroscopic data are now available 
\citep[see e.g.][]{pan2012, sutter2012SDSSdr7, ceccarelli2013, mao2016, nadathur2016BOSS} 
and recently the largest galaxy survey to date, the Dark Energy Survey (DES), has detected a 
trough and void lensing signal in a photometric survey of galaxies \citep{gruen2016, sanchez2016}, opening up 
new possibilities to exploiting the potential of voids in observations.
In the future the next generation of large galaxy surveys, such as the ESA 
Euclid mission \citep{laureijs2011, amendola2013}, are expected to provide 
a tremendous amount of new information concerning the large-scale structure 
of the Universe. The detection of gravitational lensing from medium-size voids 
in these surveys will possibly constrain the void density profiles without 
having to rely on luminous tracers like galaxies, which would require to 
model their bias \citep{izumi2013, krause2013, melchior2014, clampitt2014}. 
Nevertheless, the vast majority of available  void-finders 
\citep[see e.g.][]{Padilla_etal_2005, Platen&VandeWeygard2007, neyrinck2008, VIDE2015} 
rely on the position of dark matter particles in simulations, which cannot be directly 
compared to observables. The same finders can be adapted to use galaxies as tracers
but one will eventually need to model the tracer bias 
to compare observational results with predictions from simulations
and to fully understand properties of voids in the dark matter. 
For example, several recent works study how redshift-space distortions around 
void-centres provide constraints on cosmological parameters 
\citep{hamaus2015, hamaus2016, cai2016, chuang2016, achitouv2016, hawken2016}: all of these analyses 
are based  on the assumption that bias is linear in void environments. 
Nonetheless, a detailed study to investigate and validate this assumption is still missing.

In this paper we aim to directly determine the relation between luminous 
tracers of the large-scale structure (such as galaxies, clusters and AGN)
and their underlying matter distribution in voids to directly test the linear bias 
assumption. Thanks to state-of-the-art simulations that feature 
a full hydrodynamical treatment, the so-called {\it Magneticum Pathfinder} simulations 
\citep[Dolag et al. in prep; see also][]{hirschmann2014, saro2014, dolag2015, teklu2015, 2016arXiv160506511R}, 
we are able to perform this test with very high accuracy. The general idea 
is to run a void finder on samples of luminous objects and to extract both 
the distribution of luminous tracers and matter around void-centres, in order to 
compare them against each other.

The paper is organized as follows: in Section \ref{sims} we describe the 
simulations employed in this work, in Section \ref{meth} we present the 
commonly used bias estimators in observations and theory and discuss the 
void-finder we employ, in Section \ref{VoidMagn} we explain how we conducted our 
analysis and in Section \ref{Concl} we recap all of our result and draw our 
conclusions.

\section{The Simulations}
\label{sims}

The {\it Magneticum Pathfinder}\footnote{See http://www.magneticum.org.}
Simulations (Dolag et al., in preparation) have already been successfully
used in a wide range of numerical studies, showing good agreement with
observational findings for the pressure profiles of the intra-cluster
medium \citep{2013A&A...550A.131P,2014ApJ...794...67M}, the predicted
Sunyaev Zeldovich signal \citep{dolag2015},
the imprint of the intergalactic medium onto the dispersion
signal of Fast Radio Bursts \citep{dolag2015FastRB},
the properties of AGN population
\citep{hirschmann2014,2015MNRAS.448.1504S,2016MNRAS.458.1013S},
the dynamical properties of massive spheroidal galaxies
\citep{2013ApJ...766...71R,2016arXiv160506511R} and for
the angular momentum properties of galaxies \citep{teklu2015}.
In this paper we use the largest cosmological volume simulated, which covers a Box
with a side length of $2688h^{{-1}}$~Mpc, simulated using $2\times4536^3$
particles \citep[for details see][]{2016MNRAS.456.2361B}. We adopted a WMAP7
\citep{Komatsu11} $\Lambda$CDM cosmology with $\sigma_8 =
0.809$, $h = 0.704$, $\Omega_\Lambda = 0.728$, $\Omega_m = 0.272$,
$\Omega_b = 0.0456$, and an initial slope for the power spectrum of $n_s =
0.963$.

The simulation is based on the parallel cosmological Tree
Particle-Mesh (PM) Smoothed-particle Hydrodynamics (SPH) code {\small
P-GADGET3} (\citealp{Springel05gad}). The code uses an
entropy-conserving formulation of SPH \citep{2002MNRAS.333..649S} and
follows the gas using a low-viscosity SPH scheme to properly track
turbulence \citep{2005MNRAS.364..753D}. Based on \citet{2004ApJ...606L..97D},
it also follows thermal conduction at 1/20th of the classical Spitzer
value \citep{1962pfig.book.....S} and allows a treatment of
radiative cooling, heating from a uniform time-dependent ultraviolet
background, and star formation with the associated feedback processes. 

We model the interstellar medium (ISM) by using a sub-resolution model
for the multiphase ISM of \citet{Springel03}. In this model, the ISM
is treated as a two-phase medium, in which clouds of cold gas form by
cooling of hot gas, and are embedded in the hot gas phase assuming
pressure equilibrium whenever gas particles are above a given threshold 
density. The hot gas within the multiphase model is heated by supernovae
and can evaporate the cold clouds. A certain fraction of massive stars (10 per
cent) is assumed to explode as supernovae type II (SNII). The released
energy by SNII ($10^{51} \, {\rm erg}$) triggers galactic winds
with a mass loading rate proportional to the star formation rate
(SFR) with a resulting wind velocity of $v_{\mathrm{wind}} = 350 \, {\rm km}/s$.
Radiative cooling rates are computed by following the same procedure
presented by \citet{Wiersma09} and include a detailed model of chemical
evolution according to \citet{Tornatore07}. Metals are produced by SNII,
by supernovae type Ia (SNIa) and by intermediate and low-mass stars in the
asymptotic giant branch (AGB). Metals and energy are released by stars of
different masses, initially distributed according to a Chabrier initial
mass function \citep[IMF;][]{Chabrier03}.

Most importantly, {\it Magneticum Pathfinder} simulations include 
prescriptions for the growth of black holes and the feedback from 
active galactic nuclei (AGN) based on the model of \citet{Springel05a} 
and \citet{DiMatteo05}. Here, the accretion onto black holes and the 
associated feedback adopt a sub-resolution model. Black holes are 
represented by collisionless ``sink particles'', which can grow in mass 
by either accreting gas from their environments, or merging with other 
black holes. The radiated luminosity of the AGN in this model is related 
to the black hole accretion.
In addition, we incorporate the feedback prescription of \citet{Fabjan10};
namely, we account for a transition from a quasar- to a radio-mode feedback
whenever the accretion rate $\dot{M}_{\rm BH}$ (where $M_{\rm BH}$ is the black-hole mass) 
is low. We introduced some more technical
modifications of the original implementation, for which readers can find
details in \citet{hirschmann2014}, where we also demonstrate that the 
bulk properties of the AGN population within the simulation
are quite similar to the observed AGN properties.

We use the {\sc SUBFIND} algorithm \citep{2001MNRAS.328..726S,
2009MNRAS.399..497D} to define halo and sub-halo properties. {\sc SUBFIND}
identifies sub-structures as locally overdense, gravitationally bound groups
of particles. Starting with a main halo identified through 
the Friends-of-Friends (FoF)
algorithm with a linking length of 0.16 times the mean 
inter-particle separation, a local 
density is estimated for each particle via adaptive
kernel estimation, using a prescribed number of smoothing neighbours.
Starting from isolated density peaks, additional particles are added in
sequence of decreasing density. Whenever a saddle point in the global
density field is reached that connects two disjoint overdense regions, the
smaller structure is treated as a sub-structure candidate, and the two
regions are then merged. All sub-structure candidates are subject to an
iterative unbinding procedure with a tree-based calculation of the
potential. These structures can then be associated with galaxies, and their
integrated properties (such as stellar mass, $M_*$) can then be calculated. 
For the main haloes identified by the FoF algorithm, 
the virial radius is calculated using a density contrast based on the 
top-hat model \citep{Eke96}. For comparison with observations
we additionally use over-density with respect to 500 times the critical
density to define $M_{500c}$, which is the mass we will refer to 
as cluster mass in this paper. We refer to clusters as main haloes
with $M_{500c} > 10^{13} \, M_{\sun}/h$.

For our analysis we make use of the galaxy, cluster and AGN samples extracted 
from the simulation at redshift $z=0.14$ with the criteria explained here above. 
In Table \ref{tab:tracers} we summarise some properties of the samples which 
are relevant for our work.

\begin{table}
	\centering
	\caption{Properties of the galaxy, cluster and AGN populations
	extracted from the {\it{Magneticum}} simulations. We report the minimum mass of
        the object included, $M_{\rm min}$, in terms of stellar masses $M_{*}$ for the galaxies,
        $M_{\rm 500c}$ for clusters and $M_{\rm BH}$ for AGNs, as well as the number of
        tracers $N_t$ and of identified voids $N_{\rm v}$}
	\label{tab:tracers}
	\begin{tabular}{l|lcr} 
		Tracers  & $M_{\rm min} [M_{\sun}/h]$    & $N_t$          &  $N_{\rm v}$        \\
		\hline
                Galaxies &  $M_{*} = 4 \times \, 10^{8}$   & $9.5 \times \, 10^6$ &  $36430$      \\
                Clusters &  $M_{500c}=1 \times 10^{13}$         & $2.6 \times \, 10^6$ &  $16970$      \\  
                         &  $M_{500c}=5 \times \, 10^{13}$ & $3.5 \times \, 10^5$ &  $3125$       \\
                         &  $M_{500c}=1 \times 10^{14}$         & $1.0 \times \, 10^5$ &  $1053$       \\
                AGNs     &  $M_{\rm BH} = 4 \times \, 10^6$& $5.3 \times \, 10^6$ &  $26265$      \\
		\hline
	\end{tabular}
\end{table}

\section{Methodology}
\label{meth}
In this Section we will briefly recap how the tracer bias
is defined in observations and theory. We will also present
the void finder we used and summarize some properties
of void profiles that are relevant for this work.

\subsection{Correlation functions and bias estimation}
\label{sec:corr}
The tracer correlation function is a measure of the degree 
of clustering of the tracer itself. Being $\mathrm{d}^2 P$ the
probability that a tracer $A$ in the volume $\mathrm{d}V_A$
and another tracer $B$ in the volume $\mathrm{d}V_{B}$ are separated by a
distance $r$, the spatial two-point 
correlation function, $\xi_{AB}(r)$, is defined as the deviation of such 
probability from that expected from a random 
distribution of tracers:
\begin{equation}
\mathrm{d}^2 P = \langle n_A \rangle \langle n_B \rangle [1+\xi_{AB}(r)]\mathrm{d}V_{A}\mathrm{d}V_{B}
\end{equation}
where
$\langle n_A \rangle$ and $\langle n_A \rangle$ are the mean densities of the tracers
\citep{peebles1980}.
When we compare tracers of the same population we refer to $\xi$ as the
auto-correlation function while, if we compare two kind of tracers, we refer to $\xi$ 
as cross-correlation function.

In general there is no reason to assume that the distribution of baryons 
in the Universe traces exactly the distribution of mass. 
In fact, on small scales, galaxy formation involves many dissipative 
processes such as the radiative cooling 
of hot gas, so the efficiency of galaxy formation is 
related to how deep the potential wells created by haloes were; hence, on small scales,
bias between matter and tracers is a complicated function of space and time. 
By looking the distribution of tracers on very large scales, 
we can only observe the most luminous galaxies which are hosted by the most
massive haloes \citep{kaiser1984} i.e. by the highest peaks in the density-field.
Thereby, in the latter regime, tracers still do not perfectly mirror the same 
distribution as matter, but, since the density fluctuations are small,
the relation between matter and luminous tracers result in a constant 
offset in the clustering  amplitude, the {\it linear bias}, which, in terms of spatial 
correlations, can be written as:

\begin{equation}
  \label{eq:bias-crossC}
b = \xi_{tm}/\xi_{mm} \, ,
\end{equation}
or 

\begin{equation}
  \label{eq:bias-autoC}
  b= \sqrt{\xi_{tt}/\xi_{mm}}
\end{equation} 
where 
$\xi_{tm}$ is the tracer-matter cross-correlation function, 
$\xi_{mm}$ is the matter auto-correlation function and
$\xi_{tt}$ is the tracer auto-correlation function (the tracers being
galaxies, clusters and AGNs for our purposes). 
We make use of these two definitions to calculate the value of
the linear bias of tracers in the {\it{Magneticum}} simulation.

\subsection{Theoretical bias}
\label{sec:th-bias}

The excursion set formalism, introduced by \citet{press1974} to
predict the number of virialised dark matter haloes in the Universe
and fully developed by \citet{Bond1991}, provides 
a neat framework to develop a simple theoretical model to calculate 
the clustering of dark matter haloes and how their spatial distribution 
is biased with respect to that of the mass. With this approach, 
\citet{mo1996} estimated the bias, $b_{\rm MW}$, as

\begin{equation}
  \label{eq:biasPS}
  b_{\rm MW} = 1 + \frac{\nu^2 -1}{\delta_{\rm crit}} \, ,
\end{equation}
where 
$\delta_{\rm crit}$ is the critical density exceeding which the collapse occurs and
$\nu \equiv \delta_{\rm crit} / \sigma(M)$ is the height of the threshold in units 
of the variance of the smoothed density distribution, $\sigma(M)$, at a 
given halo mass, $M$:

\begin{equation}
  \label{eq:sigma}
  \sigma^2(M) = \frac{1}{2 \pi} \int_0^{\infty} dk k^2 P_m (k, z) \tilde{W}^2_R(k) \, ,
\end{equation}
$P_m(k,z)$ being the matter power spectrum at redshift $z$ and $\tilde{W}_{R}(k)$ the 
Fourier transform of the top hat filter function \citep[for a review of 
these topics we refer to][]{zentner2007}.

However, the bias as expressed by eq.~\ref{eq:biasPS} fails to predict with 
high accuracy the value of the bias measured in numerical simulations. For this 
reason various corrections to eq.~\ref{eq:biasPS} have been proposed in 
order to improve the consistency with simulation results 
\citep[see e.g.][]{sheth1999, sheth_mo_tormen2001, sheth2002, seljak2004, 
tinker2010}. 
In particular, \citet{tinker2010} shows how, by calibration with a 
large set of simulations, it is possible to estimate bias to 
great accuracy. Following the results of \citet{tinker2010}, 
the linear bias $b_{\rm Tinker}$ reads as: 

\begin{equation}
  \label{eq:biasTinker}
  b_{\rm Tinker} = 1 - A  \frac{\nu^a}{\nu^a + \delta_{\rm{crit}}^a} + B  \nu^b + C  \nu^c \, .
\end{equation}
where
a, A, b, B, c, C are the calibrated parameters.

We will compute the theoretical value of the linear bias in the {\it Magneticum simulations}
using both the formula by Mo\&White and its correction
by Tinker. In order to calculate
the theoretical mean value of the bias associated to our cluster sample
we will average its value using the number of objects as function of their mass
(the cluster mass function) $\frac{{\rm d}n}{{\rm d}M}$,
i.e.:
\begin{equation}
  \label{eq:mean-bias}
  \langle b \rangle = \frac{1}{\langle n_t \rangle
} \int_{M_{\rm min}}^{M_{\rm max}} \frac{{\rm d}n}{{\rm d}M} b(M) {\rm d}M \, ,
\end{equation}
where $M_{\rm min}$ and $M_{\rm max}$ are the lowest and largest masses in the
sample, respectively.

\subsection{The void finder}
\label{finder}

The biggest criticism concerning void studies is generally related to the ambiguity 
of the void definition: there are in fact many different available finders and, 
in some circumstances, the usage of such a variety of recipes to identify voids
can lead to results almost impossible to compare \citep{Colberg_etal_2008}. 
Although the void definition can be a serious obstacle on the way to establish 
a coherent picture on void properties, previous works proved that some 
statistical properties of voids (such as their number function or profile) are 
strongly affected by galaxy bias independently of the finder in use. In 
particular, recent papers exploiting the differences between voids in a
$\Lambda$CDM cosmology and modifications of gravity \citep{Cai_etal_2014}, 
Galileon or non-local gravity \citep{Barreira_etal_2015}, or possible couplings 
between cold dark matter and dark energy \citep{pollina2015} conclude that, 
while voids identified by matter particles exhibit a clear deviation from the $\Lambda$CDM case, 
it is impossible to discriminate between models looking at the statistics 
of voids identified by haloes. While \citet{Barreira_etal_2015} connect the latter result 
with the poor statistic of the halo-sample, \citet{Cai_etal_2014} suggest that 
this feature is related to the halo bias in agreement with \citet{pollina2015}, 
where the authors verified that the poor statistic of haloes is not 
sufficient to justify the dissimilar properties displayed by voids in haloes
and voids in matter. A similar conclusion has been drawn independently by 
\citet{Nadathur2015bias} using a $\Lambda$CDM simulation. It is a quite remarkable
fact that all of these works reach the same conclusion using different void-finders, namely: 
an improved version of the finder presented in \citealt{Padilla_etal_2005} 
(employed by \citealt{Cai_etal_2014}), 
the Watershed Void Finder algorithm (\citealt{Platen2007}, 
used by \citealt{Barreira_etal_2015}), 
{\small VIDE} (\citealt{VIDE2015}, utilised by \citealt{pollina2015}) and 
a modified version of {\small ZOBOV} (\citealt{neyrinck2008}, 
employed by \citealt{Nadathur2015bias}).
So, although it has been pointed out that a dynamical approach in void-finding 
(in which there is no reliance on particle positions) can reduce the impact of shot noise
in void-identification \citep[see e.g.][]{elyiv2015}, this is not relevant for 
the present study where we look directly at bias effects, which, as clarified 
above, are visible independently of the finder in use. That said, 
it is crucial to be as clear as possible in the description of the void 
finder and of a possible selection applied on top of the void catalogue to ensure 
that conclusions and results attained are plausible and can be reproduced by 
other parties. In the following lines we present the finder we employed for our analysis.

We make use of the publicly available void finder {\small VIDE} \citep[][Void IDentification and Examination toolkit]{VIDE2015} 
to identify voids. {\small VIDE} is a wrapper for {\small ZOBOV} \citep[ZOne][ZOnes Bordering On Voidness]{neyrinck2008},
an algorithm that identifies depressions in the density distribution of a set of points 
and merges them in voids with a watershed transform. Here  we 
provide a summary of how {\small ZOBOV} works, outlining the basic points of the 
procedure and we refer to the original {\small ZOBOV} paper \citep{neyrinck2008} 
and to the {\small VIDE} paper \citep{VIDE2015} for a more detailed discussion.

The void-finding technique in {\small ZOBOV} has four fundamental steps:
\begin{enumerate}
\item{{\it Voronoi tessellation} - a cell is associated with each particle $p$ following 
the prescription that a cell associated to $p$ is the region of 
the box which is closer to $p$ than to any other particle in the box. The reciprocal 
of the cell volume is an estimate of its density. Hence, with this first step we define the density 
field in which to look for voids;}
\item{{\it Definition of density minima} - the algorithm finds minima in the density 
field established in step ({\it i}). A density minimum is defined as a Voronoi cell 
with a density lower than all its adjacent cells;}
\item{{\it Creation of basins} - {\small ZOBOV} joins together cells of increasing 
density surrounding a density minimum until no adjacent higher-density cell is found, 
this defines {\it basins} as the union of these cells. Basins are depressions in the 
density field, i.e. they could be considered as voids themselves, but single basins 
may also arise from spurious Poisson fluctuations due to particle discreteness;}
\item{{\it Watershed transform} - basins are joined together using a watershed 
algorithm \citep[see][]{Platen2007} to form larger voids under the condition that 
the ridge between basins has a density lower than a given threshold (which is set 
to be the $20 \%$ of the mean density of the universe within the {\small VIDE} framework). 
This technique naturally builds up a hierarchy in the structure of voids, including 
sub-voids.}
\end{enumerate}

These steps are implemented by the enhanced version of the {\small ZOBOV} algorithm included in the 
{\small VIDE} toolkit. Additionally, {\small VIDE} calculates the void-centre 
as the volume-weighted barycentre, $\vec{x^c}$, of the cells included in a void:
\begin{equation}
  \label{v-center}
  \vec{x}^c = \frac {\sum\limits_{i=1}^N \vec{x}^{t}_i \cdot V^{t}_i }
      {\sum\limits_{i=1}^N V^{t}_i } \,,
\end{equation}
where $\vec{x}^{t}_i$ and $V^{t}_i$ are the positions of the $i-{\rm th}$ tracer (i.e. particle) $t$
and the volume of its associated Voronoi cell respectively and $N$ is the number of particles
included in the void. The effective radius of the void, $R_{\rm eff}$, is computed 
from the overall volume of the underdense region by assuming sphericity:

\begin{equation}
  V_{\rm void} \equiv {\sum\limits_{i=1}^N V^{t}_i }= \frac{4}{3} \pi R_{\rm eff}^3 \,.
\end{equation}
{\small VIDE} provides many catalogues in which various types 
of sample selections (as e.g. cuts on the void hierarchy or on 
the void central density) are applied on top of the original {\small ZOBOV} sample. 
Since for observations it is often undesirable to perform a selection on the void sample
due to poor statistics, we will apply no selection regarding void hierarchy 
or central density, thereby allowing also voids-in-clouds (voids in overdense environment) 
in our analysis.

The {\small ZOBOV} code was originally intended for void-finding in simulations, 
but {\small VIDE} also provides a flag for void-identification in light cones 
from observations including the survey mask into the analysis 
\citep[see the VIDE paper][for further information]{VIDE2015} and, in fact,
several catalogues of voids in spectroscopic samples are already available 
\citep{sutter2012SDSSdr7, pan2012, ceccarelli2013, nadathur2016BOSS, mao2016}.
Computationally we therefore have no problem in handling void-finding in observations. 
The issue left to address is how to incorporate the bias into our framework, since we
will need to relate properties of voids identified in the distribution of luminous tracers of the dark 
matter with voids in the dark matter distribution itself, that are usually under study in simulations. 
We will elucidate this relation in the following Sections of this paper.

\subsection{Density profile of cosmic voids}

The void density profile is one of the basic void statistics.
From previous studies it is known that the spherically averaged profile 
of voids exhibits a very simple structure \citep[see e.g.][]{hamaus2014, 
ricciardelli2013, ricciardelli2014}: voids are deeply 
under-dense in the vicinity of their centre and feature an over-dense 
compensation wall at $r \approx R_{\rm eff}$, where $r$ is the 
radial distance from the void centre. Such a density profile can be described 
with a simple fitting formula \citep{hamaus2014}:

\begin{equation}
  \label{eq:prof}
  \frac{n_{\rm vt}}{\langle n_t \rangle} -1 = \delta_c \frac{1-(r/r_s)^{\alpha}}{1+(r/R_{\rm eff})^{\beta}}\,,
\end{equation}
where $\delta_c$ is the central density contrast, $r_s$ is a scale radius 
at which the profile density, $n_{\rm vt}$, is equal to the average density of tracer $\langle n_t \rangle$; 
$\alpha$ and $\beta$ describe the inner and outer slopes of the void profile.

It is possible to show that the density profile of voids encodes the same 
information as the void-tracer cross correlation function.
In fact the radial profile of voids is nothing but a procedure by means 
of which we count tracers at a given distance from the void centre 
per units of volume, 
i.e. the cross-correlation between centres and tracers by definition
\citep[see, e.g.,][]{hamaus2015}. 
$N_t$ being the number of tracers, $N_{\rm v}$ the number of voids,
$\delta^D$ Dirac's delta function, $V$ the total volume,
$\vec{x}^c_i$ the coordinates of the centre of the $i-{\rm th}$ void,
and $\vec{x}^t_j$ the coordinates of the $j-{\rm th}$ tracer
we can show  explicitly that the radial spherically averaged 
void tracer-density profile compared to the mean tracer density of the Universe, is:

\begin{equation*}
\begin{aligned}
\frac{n_{\rm vt}(r)}{\langle n_t \rangle} = & \frac{1}{N_{\rm v}} \sum_i \frac{n^i_{\rm vt}(r)}{\langle n_t \rangle}= \\
                                 & \frac{1}{N_{\rm v}} \sum_i \frac{1}{N_t} V \sum_j \delta^D(\vec{x}^c_i -\vec{x}^t_j + \vec{r}) = \\
                                 & V \sum_{i,j} \int \frac{1}{N_{\rm v}}\delta^D (\vec{x}^c_i-\vec{x}) \frac{1}{N_t} \delta^D (\vec{x}-\vec{x}^t_j + \vec{r}) d^3 x= \\ 
                                 &\frac{1}{V} \int \frac{n_{\rm v}(\vec{x})}{\langle n_{\rm v} \rangle} \cdot \frac{n_t(\vec{x} + \vec{r})}{\langle n_t \rangle}d^3 x= 1+\xi_{\rm vt}(r)
\end{aligned}
\end{equation*}
thereby proving that:
\begin{equation}
\label{eq:prof-xi}
\frac{n_{\rm vt}(r)}{\langle n_t \rangle}-1 = \xi_{\rm vt}(r).
\end{equation}
as we wanted to show.

\section{The statistics of voids in the {\it Magneticum Pathfinder} simulations}
\label{VoidMagn}

The aim of this work is to study the distribution of matter around potentially 
observable voids, i.e. voids identified in the distribution of luminous tracers, such as galaxies, 
clusters, or AGNs. The basic idea is to run our void-finder (described in 
Section \ref{finder}) on the galaxy, cluster and AGN catalogues extracted from a large 
fully-hydro simulation (the {\it Magneticum}, see Section \ref{sims}) 
and calculate both the density-profile of dark matter and of its tracers around 
voids. A similar study has been conducted by \citet{sutter2014DMofG-voids}, although
the main purpose there was to show that voids in galaxies coincide with 
underdense regions of the dark matter distribution, which is indeed a 
crucial study to investigate potentially observable properties of voids. 
The authors also perform a void-to-void comparison for voids identified by galaxies and by 
matter particles. They conclude that it is always possible to identify a matter void in the 
vicinity of a galaxy void, although an offset between their centres is usually 
present. At that stage it has been concluded that potentially observable voids are 
indicative of the presence of an underdensity of matter in our Universe. As we 
have now evidence that the tracer bias is playing a fundamental role in 
void-analysis (see the discussion at the beginning of Section \ref{finder}), 
we need to further investigate the relation between tracers and the matter 
distribution around voids, which is the goal of the present work.

In the following, we will refer to galaxy-voids, cluster-voids or AGN-voids
to indicate voids which are defined by applying the finder on the galaxy-sample,
the cluster-sample or the AGN-sample. More generally
we will refer to tracer-voids to indicate the three of them at the same time.
\subsection{Dark matter distribution around void centres} 
{\label{matterArV}}

\begin{figure*}
\begin{center}
  \includegraphics[scale=0.5]{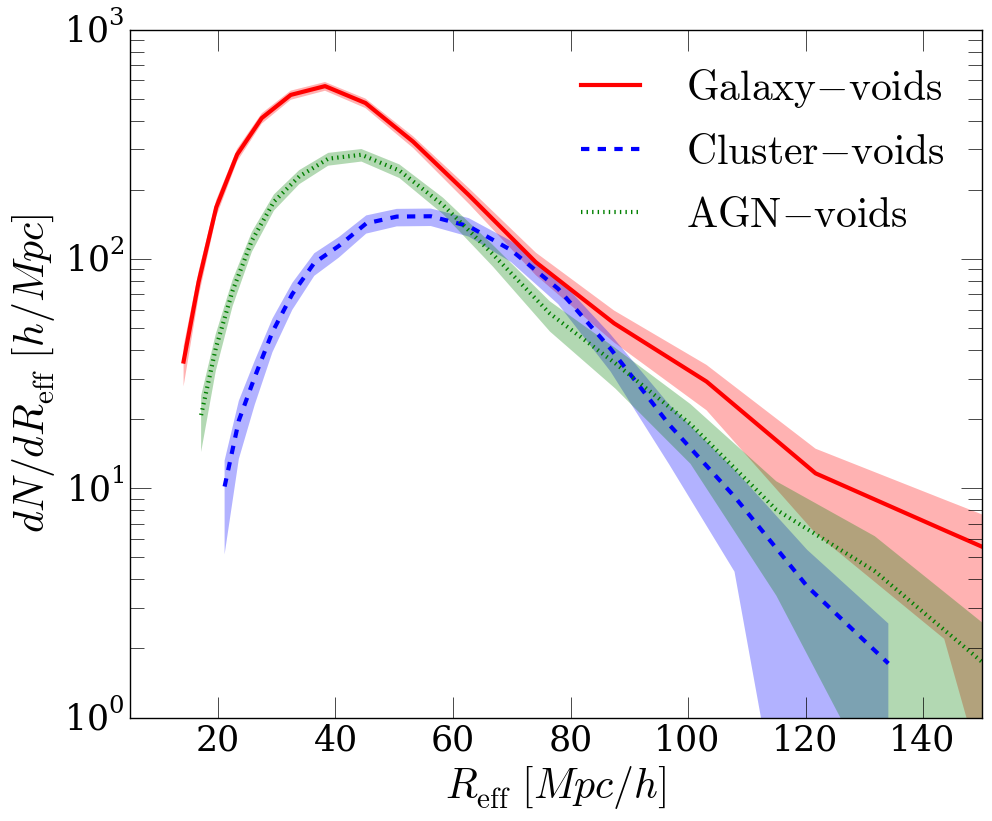}
  \caption{Abundances of voids in the {\it Magneticum} simulation. 
    Voids are identified in the distribution of galaxies (solid red line), 
    clusters (dashed blue line) and AGN (dotted green line). The shaded area represents 
    the error, calculated as Poisson uncertainty on the number 
    counts. Using galaxies as tracers of the underlying density 
    field of the Universe we are able to resolve and to find a sample 
    of voids with a typical size between $15-60 \, {\rm Mpc}/h$, twice as 
    many as the sample of AGN-voids and 4 times as many as 
    cluster-voids in the same range of size. We expect such a 
    result due to the effect of the tracer sparsity on void finding 
    \citep[see][]{sutter2014sparseS}: with a low number of 
    tracers (see Table \ref{tab:tracers}) we are not able to 
    resolve voids of small size.}
\label{fig:NumbFunc-obs-minnone}
\end{center}
\end{figure*}

To have a first overview of the void catalogues we are about to use, we look at the 
size distribution of voids which is displayed in 
Fig.~\ref{fig:NumbFunc-obs-minnone}. This figure shows the number of voids as a 
function of $R_{\rm eff}$. Voids in {\it{Magneticum}} have sizes between $15 \, 
{\rm Mpc}/h$ and $150 \, {\rm Mpc}/h$ (we refer to the largest volume simulated, 
where the box-size is $2688 \, {\rm Mpc}/h$).
As we could have expected due to the number of tracers available in each sample
(see Table \ref{tab:tracers}) and its effect on void-finding 
\citep[see][]{sutter2014sparseS} we resolve the smallest voids
in the galaxy sample, where we see twice as many voids of size within 
$15 - 60 \, {\rm Mpc}/h$ as in the AGN-sample and four 
times as many as in the cluster sample.

\begin{figure*}
  \includegraphics[width=\textwidth]{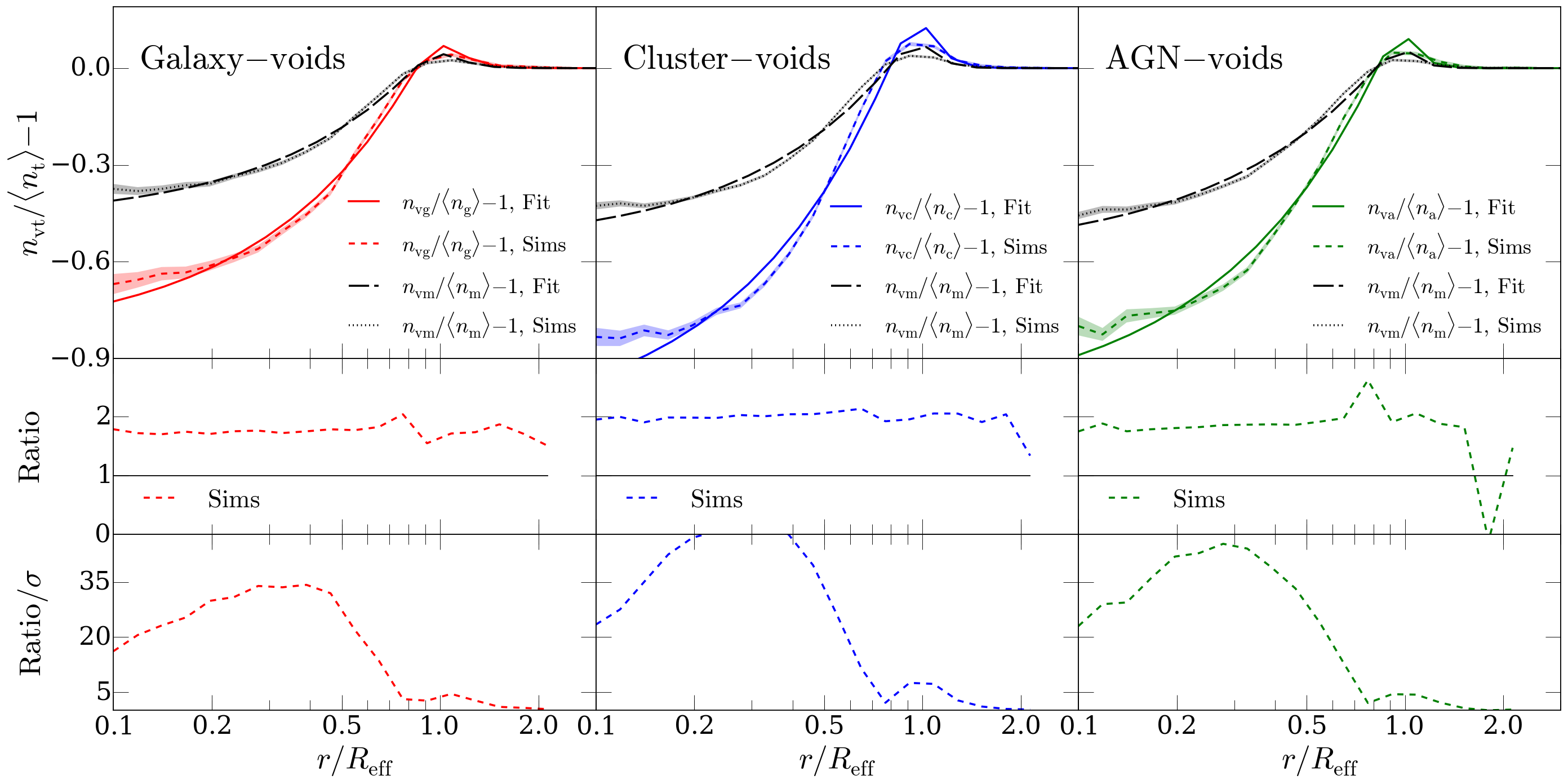}
  \caption{Top panels: measured overdensity of tracers (short dashed line) and of matter (dotted line) 
    around tracer-void centres (tracers being galaxies, clusters and AGNs from the left to the right). 
    The solid lines show the fit of the tracer-profile using eq. \ref{eq:prof}. 
    The same formula can be used to fit the matter-profiles (long dashed lines).
    The shaded areas are the uncertainty computed as the
    standard deviation from the mean profiles. In the mid panels we plot the ratios
    between tracer-profiles and matter-profiles around tracer-void centres, which look fairly constant.
    In the bottom panels we display the signal-to-noise ratios of the mid-panels. 
    As the values of the measured profiles encounter zero, the signal-to-noise drops dramatically.
    These profiles are obtained 
    by stacking voids with $80 \, {\rm Mpc}/h < R_{\rm eff} < 90 \,{\rm Mpc}/h$.}
  \label{fig:gal-mat-minnone}
\end{figure*}

After this preliminary check, we can look at the distribution of matter 
around void-centres identified by the tracers. 
In the top panel of Fig.~\ref{fig:gal-mat-minnone} we show the stacked density profiles
(i.e. the average density profile of voids of similar size) for tracer-voids with 
$80 \, {\rm Mpc}/h< R_{\rm eff} < 90 \, {\rm Mpc}/h$
(the tracers being 
galaxies, clusters and AGNs from the left to the right). Each profile is calculated 
by counting objects (tracers or dark matter particles) in the volume of spherical shells; 
the distances from void-centres are expressed in units of $R_{\rm eff}$ and the 
profile density, $n_{\rm vt}$, is expressed in terms of the mean density of tracers
in the Universe $\langle n_t \rangle$. The errors are calculated as the standard deviation
from the average density profile. We use full catalogues (for matter and tracers) without
applying any sub-sampling in order to reduce as much as possible the impact of noise
caused by sparsity.
In the top panels of Fig. \ref{fig:gal-mat-minnone}, the short dashed lines
show the measured tracer-density profiles of tracer-voids in 
{\it{Magneticum}} simulations and the solid lines are their fits computed 
with the formula given in eq. \ref{eq:prof}: as expected the fitting 
formula describes correctly the tracers' distribution around tracer-voids 
\citep{sutter2014sparseS}. The dotted lines represent the 
matter distribution around tracer-void centres, and the long-dashed lines are their 
fits again with eq. \ref{eq:prof}. The formula by \citet{hamaus2014} describes 
correctly the matter 
distribution around voids defined in tracers, too. The discrepancy in the inner
regions is due to some residual sparsity effect, which becomes more important
in the vicinity of void-centres. 
The fact that eq.~\ref{eq:prof} describes correctly also the dark matter
underdensities around tracer-voids is a first 
interesting result; 
in fact, although eq.~\ref{eq:prof} has been already successfully tested both
on matter voids and galaxy voids separately, in this particular case we are not defining voids
in the matter itself: the void-finder is run only on top of observable tracers and we then
look at the matter distribution around these potentially observable voids. So the profiles of 
voids are always self-similar and describable by eq.~\ref{eq:prof} although the
finder is not directly run on the particles with which the profile is computed. Furthermore,
in principle, once a relation between the tracer-density 
profile and the matter-density profile is established, we can link the 
latter to a potentially observable void-profile, therefore opening up the 
possibility of testing this finding with observations of voids where we can 
use the relative bias between tracers to calibrate this feature.
Going back to the top panels of Fig.~\ref{fig:gal-mat-minnone}, we observe 
that, as expected from theory and previous works \citep{sutter2014DMofG-voids}, 
the tracer distributions around tracer-voids show a steeper profile when 
compared to the matter: in the vicinity of the void-centres we measure a 
larger matter density than tracer density, while on the edge of voids 
(i.e around $r \approx R_{\rm eff}$) the tracer density is higher than
the matter density.

In the middle panels of Fig.~\ref{fig:gal-mat-minnone}, we display 
the ratio between the measured matter-profiles and tracer-profiles 
(dashed line). Although all ratios look fairly constant, there is a 
large signal-to-noise drop at  $r \approx 0.75 R_{\rm eff}$, 
i.e. where profiles have overdensities close to zero. This 
is shown clearly in the bottom panels of Fig. 
\ref{fig:gal-mat-minnone}, in which the signal-to-noise ratio 
(where the noise is computed using error propagation, starting with the error on the
density profiles)
for the mid-panels is displayed.
The fact that the ratios between tracer profiles and matter profiles
look fairly constant is very promising, but it can be too naive
to trust the values given by the profile-ratios as indicators of
the matter-tracer relation considering the large signal-to-noise
drop just discussed. 

\begin{figure*}
  \includegraphics[width=\textwidth]{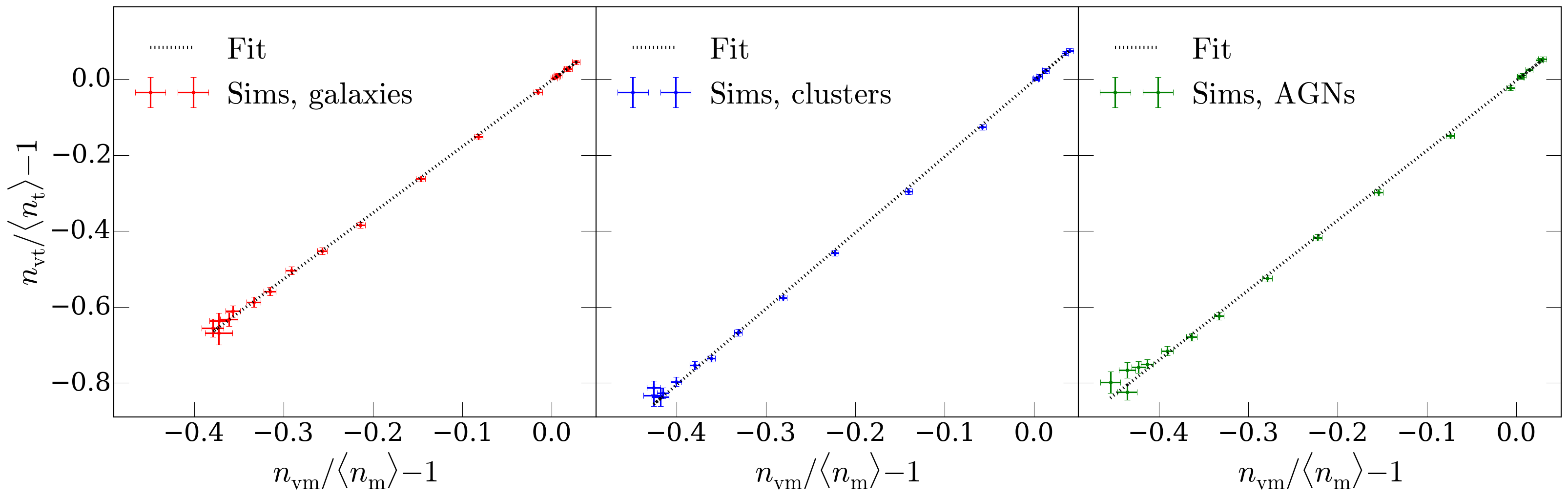}
  \caption{Overdensity of luminous tracers around voids as a function of the matter overdensity
    around tracer-void centres (both taken from Fig.~\ref{fig:gal-mat-minnone}). 
    The measured points are fitted with the linear function (dotted line) from eq.~\ref{eq:lin},
    in which the offset is consistent to zero within $5 \%$ (see Table 
    \ref{tab:linearfit}). The error bars show the standard deviation from the mean profiles. 
    From left to right the tracers are galaxies, clusters and AGNs.}
  \label{fig:mat-vs-obs}
\end{figure*}

Another way to look at the dependence 
between matter and tracer distributions around voids is to plot one 
as function of the other. In Fig.~\ref{fig:mat-vs-obs}
 we show the matter distribution around tracer-voids as 
a function of tracer distributions where the measured points are 
displayed as points with error-bars in both directions. We can fit these points with a simple linear 
function (dotted line):
\begin{equation} \label{eq:lin}
  \frac{n_{\rm vt}}{\langle n_t \rangle} - 1= b_{\rm slope} \cdot \left(\frac{n_{vm}}{\langle n_m \rangle} - 1\right) + c_{\rm offset} \, ,
\end{equation}
where
$n_{\rm vt}$ is the measured tracer-density profile around tracer-voids,
$\langle n_t \rangle$ is the mean tracer density,
$n_{vm}$ is the measured matter-density profile around tracer-voids,
$\langle n_m \rangle$ is the mean matter density of the Universe, and
$b_{\rm slope}$ and $c_{\rm offset}$ are the two free parameters of the linear 
fit, i.e. the slope and the offset, respectively. We find that the 
value of $c_{\rm offset}$ is always consistent with zero within 
$5 \%$ (except, due to sparsity, for small
voids with $20 {\rm Mpc}/h < R_{\rm eff} < 30 {\rm Mpc}/h$, in which 
$c_{\rm offset} \approx 0$ only within the $10 \%$, see Table \ref{tab:linearfit}). 
Therefore, $b_{\rm slope}$ provides a single value which fully describes the relation 
between matter and tracer 
distributions: hence, we expect $b_{\rm slope}$ to be related to the linear bias. 
In the following we will show that $b_{\rm slope}$ in fact coincides with the 
linear bias,  if sufficiently large voids are considered. 
This result suggests two main consequences: not only can we link the
tracer-profiles and matter-profile of voids using the linear bias, 
but also we can think of $b_{\rm slope}$ as a novel way to measure the bias. 
We will discuss this latter possibility in Section~\ref{comp-bias}.

\begin{table*}
	\centering
	\caption{Values of fit-parameters in eq.~\ref{eq:lin} for each tracer and void-size. We do not resolve enough cluster-voids
        with $20 \, {\rm Mpc}/h < R_{\rm eff} < 30 \, {\rm Mpc}/h $ to perform our analysis, hence we can not report the values of the parameters
        in that case.}
        \label{tab:linearfit}
        \begin{tabular}{c|cc|cc|cc} 
		\hline
		Voids & \multicolumn{2}{c}{Galaxies}  & \multicolumn{2}{c}{Clusters ($M_{500c} \geq 10^{13} M_{\sun}/h$)} &   \multicolumn{2}{c}{AGNs}  \\ 
		\hline
                \hline
                Bin in void size & $b_{\rm slope}$ & $c_{\rm offset} $ & $b_{\rm slope}$ & $c_{\rm offset} $ & $b_{\rm slope}$ & $c_{\rm offset}$ \\
                \hline
                \hline
		$20 \, {\rm Mpc}/h < R_{\rm eff} < 30 \, {\rm Mpc}/h $   &$ 2.164 \pm 0.061$ &$ -0.098 \pm 0.012$&$  -              $&$ -                $&$ 2.395 \pm 0.107$&$ -0.086 \pm 0.020 $ \\
		$30 \, {\rm Mpc}/h < R_{\rm eff} < 40 \, {\rm Mpc}/h $   &$ 2.046 \pm 0.026$ &$ -0.057 \pm 0.004$&$ 2.415 \pm 0.046 $&$ -0.041 \pm 0.005 $&$ 2.305 \pm 0.052$&$ -0.070 \pm 0.007 $ \\
		$40 \, {\rm Mpc}/h < R_{\rm eff} < 50 \, {\rm Mpc}/h $   &$ 1.890 \pm 0.014$ &$ -0.023 \pm 0.003$&$ 2.259 \pm 0.027 $&$ -0.020 \pm 0.003 $&$ 2.125 \pm 0.026$&$ -0.030 \pm 0.004 $ \\
		$50 \, {\rm Mpc}/h < R_{\rm eff} < 60 \, {\rm Mpc}/h $   &$ 1.800 \pm 0.012$ &$ -0.011 \pm 0.003$&$ 2.144 \pm 0.016 $&$ -0.011 \pm 0.002 $&$ 2.006 \pm 0.021$&$ -0.021 \pm 0.003 $ \\
                $60 \, {\rm Mpc}/h < R_{\rm eff} < 70 \, {\rm Mpc}/h $   &$ 1.751 \pm 0.011$ &$ -0.007 \pm 0.002$&$ 2.089 \pm 0.011 $&$ -0.007 \pm 0.001 $&$ 1.925 \pm 0.014$&$ -0.007 \pm 0.002 $ \\
                $70 \, {\rm Mpc}/h < R_{\rm eff} < 80 \, {\rm Mpc}/h $   &$ 1.738 \pm 0.008$ &$ -0.005 \pm 0.002$&$ 2.030 \pm 0.010 $&$ -0.005 \pm 0.001 $&$ 1.875 \pm 0.012$&$ -0.006 \pm 0.002 $ \\
                $80 \, {\rm Mpc}/h < R_{\rm eff} < 90 \, {\rm Mpc}/h $   &$ 1.746 \pm 0.006$ &$ -0.004 \pm 0.001$&$ 2.001 \pm 0.010 $&$ -0.004 \pm 0.001 $&$ 1.840 \pm 0.011$&$ -0.005 \pm 0.001 $ \\
                $90 \, {\rm Mpc}/h < R_{\rm eff} < 100 \, {\rm Mpc}/h $  &$ 1.725 \pm 0.008$ &$ -0.003 \pm 0.001$&$ 1.972 \pm 0.015 $&$ -0.003 \pm 0.002 $&$ 1.841 \pm 0.010$&$ -0.004 \pm 0.001 $ \\
                $100 \, {\rm Mpc}/h < R_{\rm eff} < 110 \, {\rm Mpc}/h $ &$ 1.767 \pm 0.010$ &$ -0.002 \pm 0.001$&$ 1.953 \pm 0.014 $&$ -0.001 \pm 0.002 $&$ 1.852 \pm 0.014$&$ -0.002 \pm 0.001 $ \\
                $110 \, {\rm Mpc}/h < R_{\rm eff} < 120 \, {\rm Mpc}/h $ &$ 1.751 \pm 0.007$ &$ -0.002 \pm 0.001$&$ 1.908 \pm 0.017 $&$ -0.003 \pm 0.002 $&$ 1.862 \pm 0.022$&$ -0.001 \pm 0.002 $ \\
                $120 \, {\rm Mpc}/h < R_{\rm eff} < 130 \, {\rm Mpc}/h $ &$ 1.735 \pm 0.021$ &$ -0.002 \pm 0.001$&$ 1.958 \pm 0.019 $&$ -0.003 \pm 0.002 $&$ 1.892 \pm 0.020$&$ -0.001 \pm 0.001 $ \\
                $130 \, {\rm Mpc}/h < R_{\rm eff} < 150 \, {\rm Mpc}/h $ &$ 1.764 \pm 0.012$ &$ -0.004 \pm 0.001$&$ 1.951 \pm 0.065 $&$ -0.007 \pm 0.008 $&$ 1.869 \pm 0.021$&$ -0.004 \pm 0.002 $ \\
                \hline
	\end{tabular}
\end{table*}

\begin{figure*}
  \includegraphics[width=\textwidth]{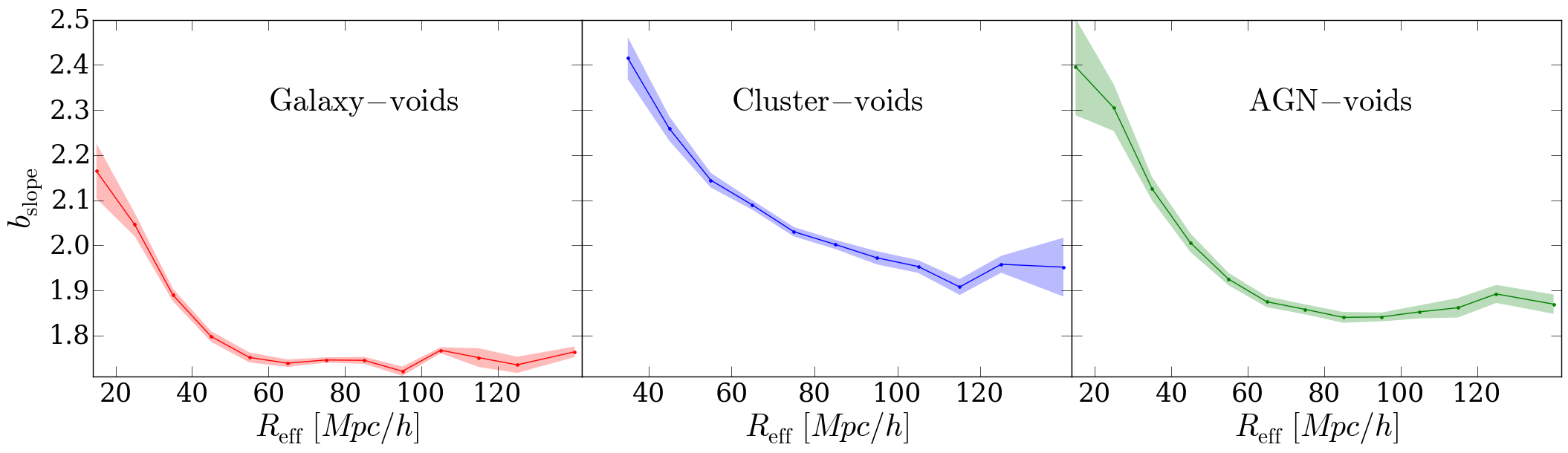}
  \caption{Value of $b_{\rm slope}$ from Fig.~\ref{fig:mat-vs-obs} for galaxies (left-panel),
    clusters (mid-panel) and AGNs (right-panel) around galaxy-voids, cluster-voids and AGN-voids 
    respectively, in various void-radius bins (i.e. as a function of void-size). The 
    shaded area represents the uncertainty, obtained from the error on the fit.
    We see an impact of void-size on the measurement of $b_{\rm slope}$, which becomes larger for small voids.}
  \label{fig:Bias-vs-R-minnone}
\end{figure*}

What we have shown so far refers to voids with an effective radius
$R_{\rm eff}$ within  
$80 \, {\rm Mpc}/h$ and $90 \, {\rm Mpc}/h$
as a guiding example,  
but we did perform our 
analysis using voids of various sizes.
We report the bins in which the analysis 
is repeated in the first column of table \ref{tab:linearfit}. 
The bins are selected such that:
\begin{enumerate}
\item{a sufficient number of voids is included in each bin so that the averaged profile is accurate enough to make 
the profile-fit and the linear-fit converge (i.e. at least $\approx 50$ voids per bin);} 
\item{the physical dimension of each single bin
is not too extended, in order to work under the hypothesis of considering voids of 
similar sizes, which is required by eq.~\ref{eq:prof};} 
\item{all void-sizes are covered.}
\end{enumerate}
We
verified that in each bin we can always fit the relation 
between matter and tracers around voids with a simple linear relation.
In Fig.~\ref{fig:Bias-vs-R-minnone}, we show the values 
for $b_{\rm slope}$ (see eq. \ref{eq:lin}) as a function 
of void-size (the tracers being, from the left, galaxies, clusters and AGNs).
We see a trend: the value of 
$b_{\rm slope}$ decreases with the increase of void-size, 
showing that small voids yield a larger bias. As the size of voids
surpasses a critical size, the value of $b_{\rm slope}$ stabilises 
asymptotically to a constant value. The critical void-size at which
$b_{\rm slope}$ becomes stable seems to be dependent on the tracer
properties (clusters, mid-panel, seem stable only starting at $R_{\rm eff} \approx 80 \, {\rm Mpc}/h$,
while galaxies and AGNs show a stable value of $b_{\rm slope}$ for roughly $R_{\rm eff} > 50 {\rm Mpc}/h$)
and on the number of tracers (i.e. on the sparsity of the sample).
The shaded areas in Fig.~\ref{fig:Bias-vs-R-minnone}  
represent the uncertainty obtained from the linear fits.

\begin{figure*}
  \includegraphics[width=\textwidth]{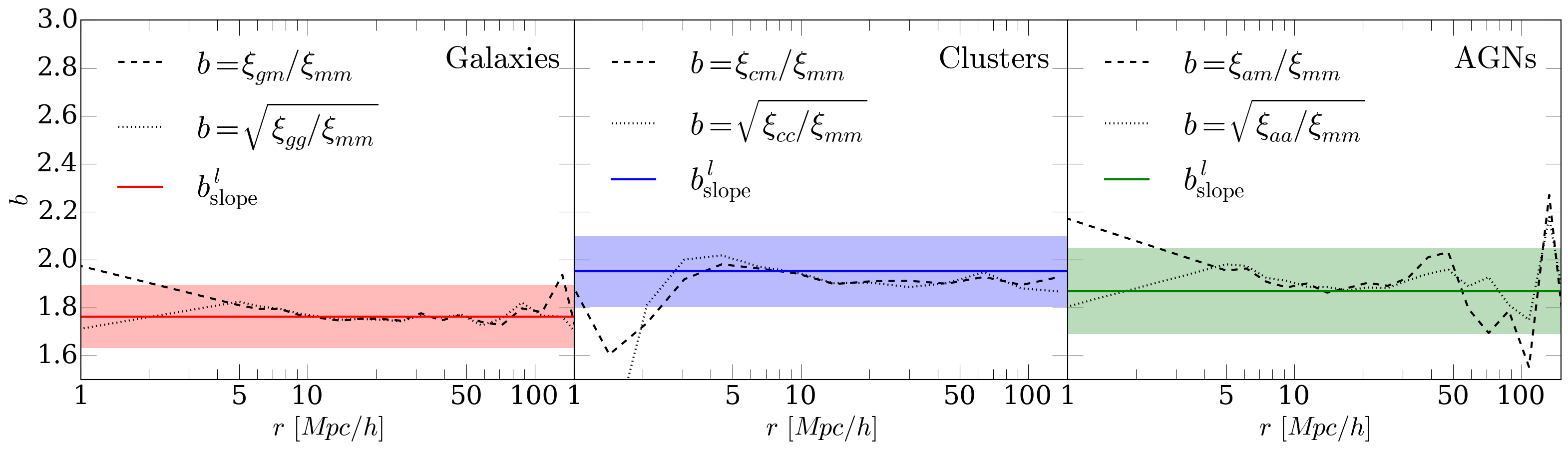}
  \caption{Comparison between different bias estimators: starting from the left panel, we plot 
    the bias of galaxies calculated as the saturated value of the slopes from 
    Fig.~\ref{fig:Bias-vs-R-minnone},
    and the usual galaxy bias estimators presented in eq.~\ref{eq:bias-crossC} (dashed 
    line) and eq.~\ref{eq:bias-autoC} (dotted line). The central and right 
    panel show the cases of clusters and AGNs in blue and green respectively. We find 
    a good consistency between $b_{\rm slope}^l$ and other bias estimators in the 
    large-scale limit. The shaded area is the error, computed as the standard deviation from the
    mean value of $b_{\rm slope}$ from all void sizes.}
  \label{fig:gal-slope-vs-bias-minnone}
\end{figure*}

\subsection{Linear bias}
\label{comp-bias}

To demonstrate convincingly that $b_{\rm slope}$ from our fit with eq. \ref{eq:lin}
is an indicator of the tracer bias, we compare its values with the most 
commonly used bias estimators (discussed in Section \ref{sec:th-bias}). 
In Fig.~\ref{fig:gal-slope-vs-bias-minnone} we show the bias computed with 
eqs.~\ref{eq:bias-crossC} (dashed lines) and \ref{eq:bias-autoC} (dotted lines) 
and 
$b_{\rm slope}^l$ (solid lines), defined as the value of $b_{\rm slope}$
calculated in the bin that includes the largest voids of each sample (i.e. $130 \, {\rm Mpc}/h < R_{\rm eff} < 150 \, {\rm Mpc}/h $):
the aim is to confront the asymptotic measured value of 
$b_{\rm slope}$ (see Fig. \ref{fig:Bias-vs-R-minnone}) with the linear bias. The shaded
areas show the error on the mean value of $b_{\rm slope}$ from all void sizes.

As we can see in Fig. \ref{fig:gal-slope-vs-bias-minnone}, for all tracers
(from the left to the right: galaxies, clusters and AGNs) $b_{\rm slope}^l$ agrees well 
with the bias calculated by eqs. \ref{eq:bias-crossC} and \ref{eq:bias-autoC} in the 
large-scale limit, but deviates on small scales.
This result confirms that the linear bias gives a good 
description of the relation between luminous tracers and matter around voids,
as long as sufficiently large voids are under study.
Hence, computing the slope in eq. \ref{eq:lin} with a simple 
linear fit, as described in Section \ref{matterArV}, provides another technique 
to estimate the linear bias in simulations, if sufficiently 
large voids are analysed. Such a technique, in principle, 
allows to extend the measurement of linear bias to 
smaller scales inside voids. A caveat is the large uncertainty: 
close to void-centres Poisson noise increases and we should carefully
consider how to further test this extension.  

\begin{figure*}
  \includegraphics[width=\textwidth]{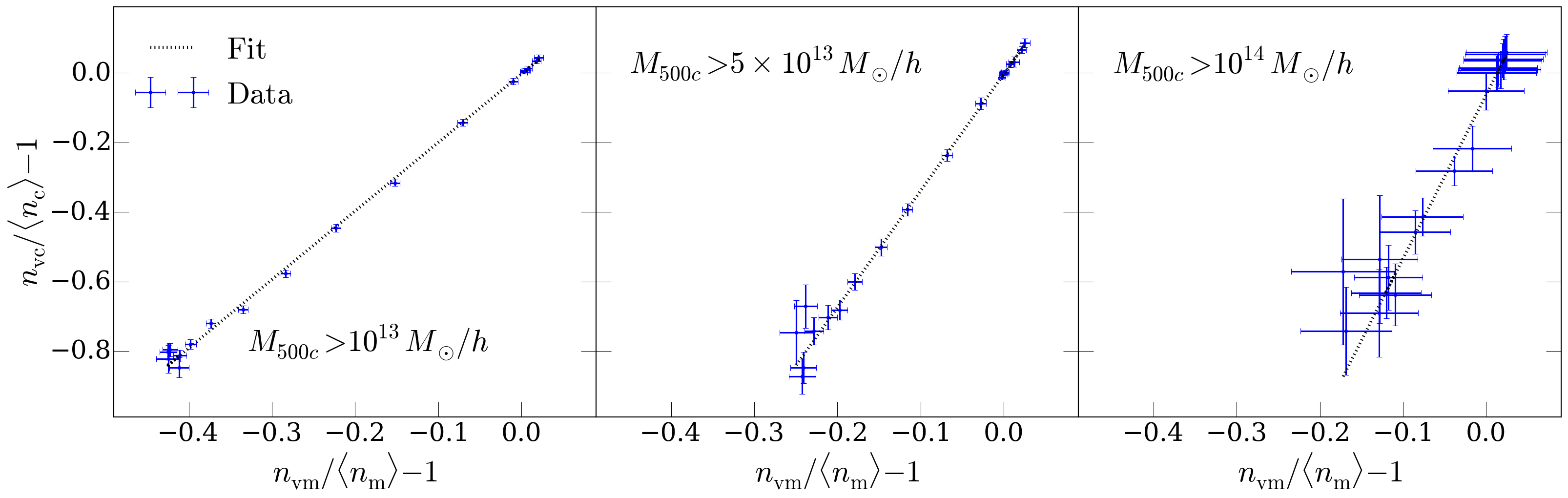}
  \caption{Distribution of clusters around cluster-void centres as a function of the 
    matter-distribution around cluster-voids measured after applying various mass 
    cuts: from the left to the right $M_{500c} > 10^{13} \, M_{\sun}/h$ (full sample), 
    $M_{500c} > 5 \times \, 10^{13} \, M_{\sun}/h$ and $M_{500c} > 10^{14} \, M_{\sun}/h$. 
    The measured 
    points (blue) are fitted with a linear function (dotted line) presented in 
    eq.~\ref{eq:lin}. The slope increases from the left to the right panel
    as expected due to the increasing mass of the objects included in the analysis. We are
    showing the plot for the largest voids included in each sample, i.e. from left 
    to right for voids with size $130-140 \, {\rm Mpc}/h$, $170-200 \, {\rm Mpc}/h$ 
    and $220-290 \, {\rm Mpc}/h$}
  \label{fig:mat-vs-obs-Mcuts}
\end{figure*}

To additionally examine the consistency of our procedure to calculate the bias, 
we verify how $b_{\rm slope}$ changes after imposing 
various mass-cuts on the cluster sample. Namely, we demand the
cluster mass to be $M_{500c} \geq 5 \times \, 10^{13} \, M_{\sun}/h$ and 
$M_{500c} \geq 10^{14} \, M_{\sun}/h$. 
To be conservative we rerun the void 
finding algorithm after applying each selection cut on top of the 
cluster-sample and we repeat the stacking procedure. By imposing these 
cuts we include a smaller number of clusters in the analysis (see 
Table~\ref{tab:tracers}), which implies that we are not able to resolve 
the smallest voids due to the sparsity of the tracers 
\citep[see e.g.][]{sutter2014sparseS}. In order to include a sufficient 
number of voids we need to modify the binning employed 
in the stacking procedure. In fact, since we are not able to resolve the 
smallest voids, it is necessary to remove, shift, or enlarge some of the bins.
Being able to resolve only very large voids in the sample of clusters 
with $M_{500c} \geq 10^{14} \, M_{\sun}/h$, 
we do not expect to find a high accuracy result; however we 
aim to find at least a qualitative indication that the bias that we 
measure using $b_{\rm slope}^l$ increases as expected in this case.

\begin{figure*}
  \includegraphics[width=\textwidth]{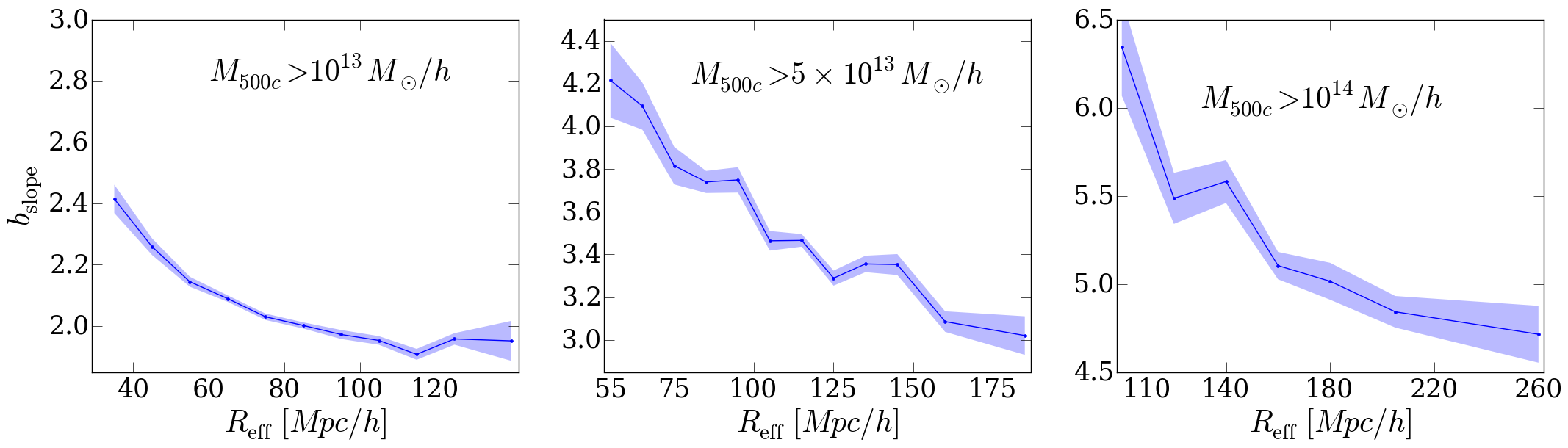}
  \caption{Values of $b_{\rm slope}$ after applying different mass cuts on the cluster sample  
  	as a function of void-size. The 
        shaded area represents the uncertainty, obtained from the error on the fit. 
        In agreement with previous findings
        we see an impact of void-size on the measurement of $b_{\rm slope}$, which becomes 
        largest for small voids. In the cases in which 
        a mass cut is applied (central and right panel, 
        $M_{500c} \geq 5 \times \, 10^{13} \, M_{\sun}/h$ 
        and $M_{500c} \geq 10^{14} \, M_{\sun}/h$)
        we observe that $b_{\rm slope}$ decreases as 
        the void-size increases, although in a noisy manner. It is also not clear whether 
        the convergence to the value of the linear bias is reached as in the full cluster 
        sample (left panel).}
  \label{fig:allClCutBias-vs-Reff}
\end{figure*}

In Fig.~\ref{fig:mat-vs-obs-Mcuts} we show the distribution of clusters 
around cluster-void centres as a function of their matter distribution (in 
analogy to Fig.~\ref{fig:mat-vs-obs}); we are displaying the relation 
for the largest voids in each sample, i.e. for voids with a $R_{\rm eff}$ 
in the bin-size range (from the left to the right), of $130-140 \, 
{\rm Mpc}/h$, $170-200 \, {\rm Mpc}/h$ and $220-290 \, {\rm Mpc}/h$.
After applying the mass-cuts we are still able to fit the matter-tracer 
relation with a simple linear dependency (i.e. using eq.~\ref{eq:lin}) 
where the offset value is consistent with zero. The exclusion of low-mass 
clusters from our main samples increases the noise on our measurement, 
which is now not as well determined as in Fig.~\ref{fig:mat-vs-obs}: the error 
bars are larger than in the full cluster-sample case and the simulation 
points are sometime further away from the fit (dotted line). However, we 
can clearly see that $b_{\rm slope}$ increases from the left to the 
right panels of Fig.~\ref{fig:mat-vs-obs-Mcuts}, following an expected trend, 
since we are imposing a larger and larger threshold on 
the mass-cut (therefore including only objects with higher and higher 
bias). 

As previously done with the other tracers under study, we performed our analysis on
voids of various size. We show how the value of $b_{\rm slope}$
changes as a function of void-size in Fig.~\ref{fig:allClCutBias-vs-Reff}:
going from the left to the right we display the curve for clusters
with $M_{500c} \geq 10^{13} \, M_{\sun}/h$ (same as Fig.~\ref{fig:gal-slope-vs-bias-minnone}, central panel,
reported here for comparison), $M_{500c} \geq 5 \times \, 10^{13} \, M_{\sun}/h$ and
$M_{500c} \geq 10^{14} \, M_{\sun}/h$. For what concerns the cases in
which a mass cut is applied (central and right panels) we see, in agreement
with our previous findings,
that small voids yield a higher value of $b_{\rm slope}$, although the trend is not 
as clearly saturating 
as in the full-cluster sample (left panel).  
For the samples in which clusters have a mass $M_{500c} \geq 5 \times \, 10^{13} \, M_{\sun}/h$ and
$M_{500c} \geq 10^{14} \, M_{\sun}/h$ we were also expecting an increasing critical void-size 
at which $b_{\rm slope}$ converges to a constant value, due both to the inclusion of highly 
biased tracers and due to their increased sparsity. However,
in this cases $b_{\rm slope}$ does not converge to a constant 
value. It is therefore
not clear whether we resolve voids large enough to reach the convergence 
value in the central and right panels.

\begin{figure*}
  \includegraphics[width=\textwidth]{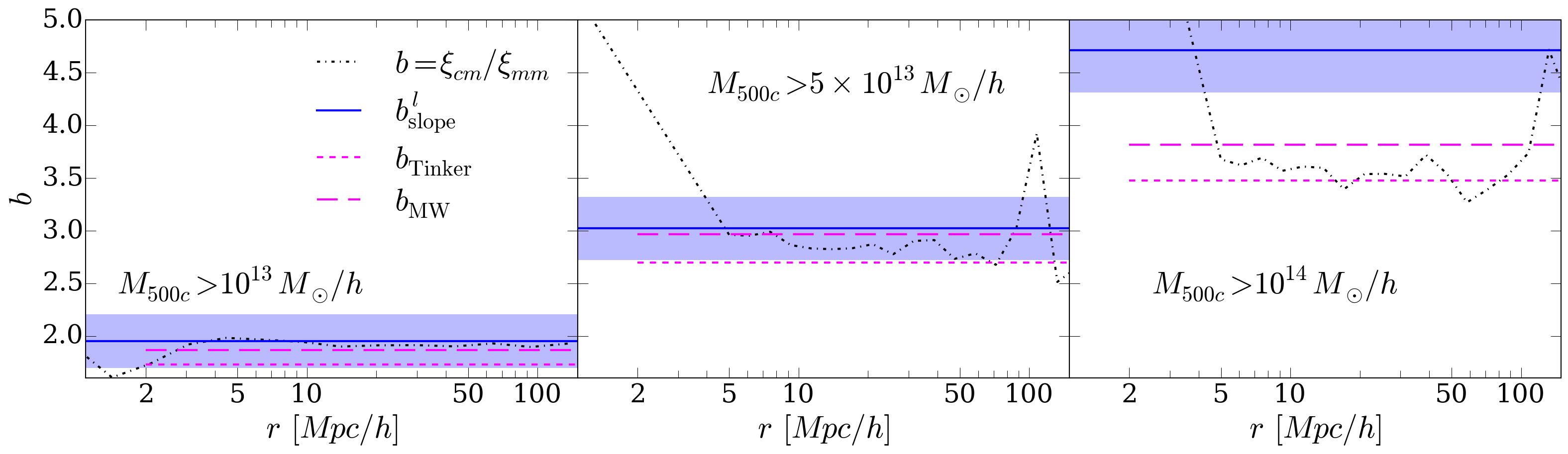}
  \caption{Comparison between different bias estimators and theory after applying various 
    mass-cuts: we plot the bias of clusters calculated as $b_{\rm slope}^l$ 
    (solid line), and the classical bias estimator (eq.~\ref{eq:bias-crossC}, 
    dashed-dotted line). The value predicted by eq.~\ref{eq:biasTinker} and eq.~\ref{eq:biasPS}
    is shown by the dashed line and the long dashed line respectively. The shaded areas 
    represent the standard deviation 
    from the mean values of $b_{\rm slope}$.}
  \label{fig:cl-slope-vs-bias-mincuts}
\end{figure*}

We can now compare 
$b_{\rm slope}^l$ with the bias calculated using eqs.~\ref{eq:bias-crossC}, 
\ref{eq:bias-autoC}. Since we are only considering clusters we can now
include theory predictions to our plot, using the Mo\&White formula 
(eq.~\ref{eq:biasPS}) and its extension  by Tinker (eq.~\ref{eq:biasTinker}).
To estimate the mean value of $b_{\rm MW}$ and $b_{\rm Tinker}$ we use as weight the 
theoretical mass function calculated with \citet{press1974} and the \citet{tinker2010}
respectively.

In Fig.~\ref{fig:cl-slope-vs-bias-mincuts} we plot the values of linear bias calculated 
with all of these methods. Namely:
\begin{itemize}
  \item{$b_{\rm slope}^l$ (solid line, the shaded area represents the error on 
    the mean value of $b_{\rm slope}$ for voids of various sizes);}
  \item{$b$ from eq.~\ref{eq:bias-crossC} (dashed-dotted line);}
  \item{$b_{\rm MW}$ from eq.~\ref{eq:biasPS} (long dashed line);}
  \item{$b_{\rm Tinker}$ from eq.~\ref{eq:biasTinker} (dashed line).}
\end{itemize}

For the most numerous cluster sample under study ($M_{500c} \geq 10^{13} M_{\sun}/h$,
left panel), we are showing the same plot as Fig.~\ref{fig:gal-slope-vs-bias-minnone} 
(central panel) for the comparison. 
In this case and for
$M_{500c} \geq 5 \times \,  10^{13} \, M_{\sun}/h$ (central panel) we find a good agreement of $b_{\rm slope}^l$,
both with theory and with theoretical bias computed with eqs.~\ref{eq:bias-crossC} 
and \ref{eq:bias-autoC}. Values predicted by other bias estimators
are within the uncertainty. As we commented before, the agreement is 
remarkable on large scales while on small scales the traditional bias estimators deviate
from $b_{\rm slope}^l$.
For what concerns the analysis in the sample which includes only clusters 
with $M_{500c} > 10^{14} M_{\sun}/h$ (right panel), we see a significant
deviation of $b_{\rm slope}^l$ from other bias predictions. We expected the latter case
to be the most problematic, given the large noise 
due to the extreme mass cut applied (see Table \ref{tab:tracers}). Beside this, there
is a practical motivation for such a discrepancy: as we suspected, we did not resolve 
enough large voids to determine the convergence of $b_{\rm slope}$. We are, in fact, 
forced to include voids of a wide
range of sizes in the bin that contains the largest voids in this sample 
($220-290 \, {\rm Mpc}/h$) in order to obtain sufficiently smooth profiles and reach the
convergence in the linear fit between matter-void profile and cluster-void density profile.
This is absolutely necessary: if we would include only voids with sizes between e.g. $225-290 \, {\rm Mpc}/h$
the linear fit would not converge. Apparently, for this particular case, voids with $R_{\rm eff} \approx 220 \, {\rm Mpc}/h$
are too small to attain the convergence of $b_{\rm slope}$ to the value of the linear bias.
Ideally, if we had resolved a sufficient number of very large voids ($R_{\rm eff} > 250 \, {\rm Mpc}/h$) we would recover the value of 
the linear bias also in this case, but the simulation box is too small to get a sufficiently large number of voids of that size.
This is indeed a critical point
as, the only way to tell if the $b_{\rm slope}$ converged to a saturated value is by
looking at fig.~\ref{fig:Bias-vs-R-minnone} and fig.~\ref{fig:allClCutBias-vs-Reff}. 
Moreover, for practical purposes we can use the relative bias between different tracers 
for that matter, which is accessible in observations.

As we commented before, it is not clear by Fig.~\ref{fig:allClCutBias-vs-Reff} whether
$b_{\rm slope}$ converges to a constant value for the cluster samples with 
$M_{500c} \geq 5 \times \, 10^{13} \, M_{\sun}/h$  and
$M_{500c} \geq 10^{14} \, M_{\sun}/h$: looking at 
Fig.~\ref{fig:cl-slope-vs-bias-mincuts} we can conclude that the convergence is 
reached for $M_{500c} \geq 5 \times \, 10^{13} \, M_{\sun}/h$ but not for 
the most extreme mass cut. It is remarkable that, in the latter case, we 
understand why the convergence of $b_{\rm slope}^l$ can not occur. 
However, we have demonstrated that our method of calculating 
the bias with $b_{\rm slope}^l$ is quite consistent when applied on samples
with various masses which was the aim of this test.

To summarize this Section, we have shown that the relation 
between matter and matter-tracers in voids is always linear
and determined by a single number $b_{\rm slope}$. 
This result was established by directly measuring the 
distribution of matter and tracers around voids and incidentally 
validates recent work that simply assumed the bias of tracers 
to be linear in the vicinity of voids.
Furthermore we showed that, by measuring the matter profile and
the matter-tracer profiles around large voids in simulations, we can estimate the 
value of the linear bias via the slope of a simple linear fit between the two 
distributions, if sufficiently large voids are considered.

\section{Discussion and Conclusions}
\label{Concl}

With the help of a suite of state-of-the art hydro-simulations
we have investigated the stacked tracer-density profile of cosmic voids and linked
it to their underlying matter-density profile. Before discussing the implications of
our findings, we recap all major results of this work:

\begin{itemize}\itemsep2pt

\item{
      {\bf{The underlying matter-density profile of tracer-voids is well 
      described by the fitting formula presented in \citet{hamaus2014}}}. Such a formula 
      was known to describe the profile of tracer-voids and matter-voids separately, 
      but in this work we have successfully tested it on depressions of the matter-density field around
      tracer-voids, i.e. without running the void finder on dark matter particles. This result
      points out once again the degree of self-similarity of underdense regions
      in our Universe, as they can always be described by the same fitting function. 
}

\item{
      {\bf{The relation between the density of tracers and matter around voids is always linear and 
       determined by a multiplicative constant ($b_{\rm slope}$)}}. 
       This remarkably simple relation was tested using galaxy, cluster and AGN 
       samples extracted from {\it Magneticum Pathfinder}, including voids of various 
       sizes and applying different mass cuts on top of the cluster sample. The linear 
       relation between matter and tracers always stands, regardless of tracer type and 
       host-halo mass range.
}

\item{
  {{\bf The value of the multiplicative constant decreases with the increase of the size 
    of voids and asymptotes to the linear bias.}} 
    For sufficiently large voids, $b_{\rm slope}$ is shown to match the linear bias 
    extracted from the usual tracer auto-correlation, the tracer cross-correlation with 
    dark matter, and the expectations from theory, such as the bias 
    functions proposed by Mo\&White and Tinker. The critical void-size at which 
    $b_{\rm slope}$ converges to the linear 
    bias is dependent on the clustering properties of the tracers under study (and on 
    their sparsity). In fact, we find that for the full cluster sample the critical void-size 
    is around $80 \, {\rm Mpc}/h$ while for AGNs it is around $60 \, {\rm Mpc}/h$ and for 
    galaxies about $50 \, {\rm Mpc}/h$. 
    In order to eliminate the effect of sparsity, we sub-sampled all tracers
    to the density of our full cluster sample. This test reveals that the critical void size 
    at which $b_{\rm slope}$ reaches a saturated value also depends on tracer properties other
    than density, such as their bias.
    If a highly biased and very sparse population is used as 
    tracer of the density field, it can be possible that not enough large voids are available 
    and hence we can not establish the constant value to which $b_{\rm slope}$ converges.
    The large values of $b_{\rm slope}$ obtained by small voids show that they
    yield a biased result. We leave further investigations on the origin
    of this effect to future studies.
}

\end{itemize}

The correspondence between $b_{\rm slope}$ and linear bias is expected at linear order 
in the density fluctuations, because we can consider the stacked tracer-density profile 
of voids as a void-tracer cross-correlation function $\xi_{\rm vt}(r)$ (see eq.~\ref{eq:prof-xi}), 
and express it in terms of the void-matter cross-correlation function,

\begin{equation}
 \xi_{\rm vt}(r) = b\xi_{\rm vm}(r)\;, \label{eq:bias_voidC}
\end{equation}
via the linear tracer bias $b$. We find that eq.~(\ref{eq:bias_voidC})
is in principle valid for arbitrarily small values of $r$, 
as long as large enough voids are considered, in stark contrast 
to the common two-point statistics of tracers appearing in eqs.~\ref{eq:bias-crossC} 
and \ref{eq:bias-autoC}. 
For the latter, the linear bias model can break down below scales on the 
order of $\sim50 {\rm Mpc}/h$ at low redshift and is therefore not applicable
for a dominant fraction of available Fourier modes of the density field.

Furthermore, this technique can yield important advantages for 
the analysis of survey data. In order to maximize the amount of 
cosmological information contained within the common two-point 
statistics of large-scale structure one has to make use of 
sophisticated perturbation theory frameworks to consistently 
include all higher-order bias parameters 
\citep[e.g.][]{fry1993,mcdonald2009,beutler2016,sanchezA2016}. Alternatively, 
one can marginalize over the unknown free parameters of an empirical 
function that models non-linear bias in a phenomenological way. 
However, both approaches are very limited, they quickly break down 
towards smaller scales and the total information gain does not scale 
with the additional number of modes included.

As long as tracer bias remains scale-independent, as shown to be the case in 
void-tracer cross-correlations, these limitations do not apply. 
An example for such a cosmological analysis is the study of 
redshift-space distortions around voids \citep[][]
{hamaus2015, hamaus2016, cai2016, chuang2016, achitouv2016, hawken2016}, 
but the interpretation of many other observables, such as void abundance, 
void lensing, void clustering and the void ISW effect may benefit from a linear bias treatment 
as well. An exciting perspective is to look out for additional physical 
effects that induce a non-linear scale-dependent tracer bias around voids, 
and are typically neglected in standard $\Lambda$CDM. One such example is 
the effect of massive neutrinos \citep{loverde2014,Castorina2015,Carbone2016,banerjee2016}. Similar signatures can be expected in scenarios 
of modified gravity~\citep{Cai_etal_2014,Achitouv_etal_2016} or coupled dark 
energy~\citep{pollina2015}. We leave further research along 
these lines for future work.

\section*{Acknowledgments}
We are thankful to Paul Sutter, Guilhem Lavaux
and Steffen Hagstotz for useful discussions. 
KD and JW acknowledge the support of the DFG Cluster of Excellence 
``Origin and Structure of the Universe''  
and the Transregio programme TR33 
``The Dark Universe''.
MB acknowledges support from the Italian Ministry for Education, 
University and Research (MIUR)
through the SIR individual grant SIMCODE, project number RBSI14P4IH.
LM acknowledges financial contribution from the agreement ASI 
n.I/023/12/0 ``Attività relative alla fase B2/C per la missione Euclid''.
The calculations have partially been carried out on the computing facilities of 
the Computational Center for Particle and Astrophysics (C2PAP) and of the Leibniz 
Supercomputer Center (LRZ) under the project IDs pr58we, pr83li, and pr86re. 
Special thanks go to LRZ for the opportunity to run the Box0 simulation within 
the Extreme Scale-Out Phase on the new SuperMUC Haswell extension system. 
We appreciate the support from the LRZ team, especially N. Hammer, when 
carrying out the Box0 simulation.



\input{Magneticum_Voids_clean.bbl}

\label{lastpage}
\end{document}